\documentclass[twocolumn,twoside,slac_two]{revtex4}
\usepackage{graphicx}
\usepackage{fancyhdr}
\begin{document}

\title{Linear Polarization as a Probe of Gamma Ray Flaring Blazar Jets}

%

\author{M. F. Aller, H. D. Aller, P. A. Hughes}
\affiliation{University of Michigan, Ann Arbor, MI, 48109-1042, USA}

\begin{abstract}
We describe and present initial results from  a {\it Fermi} cycle 2 program
designed to monitor the behavior of the
centimeter-band linear polarization and total flux density emitted by gamma-ray-bright blazars 
during flaring. The goal of the program is to identify changes
in the magnetic field structure in the radio jet associated
with gamma-ray flaring and ultimately to test whether gamma-ray flaring is associated with the
onset of shocks in the radio jet. Light curves illustrating radio band variability patterns are shown for
sample program sources.

\end{abstract}

\maketitle 

\thispagestyle{fancy}


\section{Overview of Project}
As part of a {\it Fermi} cycle 2 program, we are obtaining monitoring observations at three
centimeter band frequencies (14.5, 8.0 and 4.8 GHz) with the  
University of Michigan 26-m radio telescope (UMRAO) of the source-integrated total flux
density and linear polarization of approximately 30 radio- and $\gamma$-ray- bright
sources currently or expected to be in $\gamma$-ray-flare phase.
The goal of our work is to identify jet conditions responsible for the generation
 of the $\gamma$-ray emission
and to test the hypothesis that
shocks play a role in the production of this emission as suggested by earlier work \cite{Jorstad01}. 
A causal connection  between activity in the
radio jet and $\gamma$-ray flaring first proposed based on EGRET data
\cite{Jor01b,Valtaoja95,Aller96} is supported by early
{\it Fermi} results. For example,
statistical studies of the sources in the 15 GHz MOJAVE VLBA sample \cite{Lister09} have found that
the members detected during the first three months of {\it Fermi} operation \cite{Abdo09}
are more core-dominated, have higher brightness temperatures and  Doppler factors, and are
in a more active radio state than non-detected MOJAVE sources \cite{Kovalev09,Sav09}.

While there has been increasing evidence suggesting that the $\gamma$-ray emission arises near the radio
core \cite{Jor01b}, a region believed to correspond to either the $\tau$=1 surface or to a standing shock \cite{Marscher09} and
not at a site near to the black hole/accretion disk, a number of important questions remain. These include the
trigger for the emission of $\gamma$-rays, the position of the emission site (upstream or
downstream of the radio core) and the nature of the emitting region itself (standing shock, 
propagating shock, or `blob' with a chaotic magnetic field where turbulence produces the very high energy
electrons). A plausible scenario for the broadband features seen as outbursts or flares is that
instabilities develop naturally within the collimated
relativistic jet outflows giving rise to shocks \cite{Hughes89}.
With the passage of a shock, there is a compression of the
magnetic field within the emitting region and an increase in the degree of
order of the magnetic field. The observational signature of such an event is a swing 
in the electric vector position angle (hereafter EVPA; an
orientation orthogonal to the magnetic field direction in a transparent source)
and an increase in the fractional linear polarization. 

\begin{table}[t]
\begin{center}
\caption{Program Sources}
\begin{tabular}{|l|l|l|l|l|}
\hline  0048$-$097 & NRAO 190   & 0917$+$449 & 1510$-$089 & 2022$-$077    \\
\hline  0109$+$224 & 0454$-$234 & 1127$-$145 & 1633$+$382 & BL Lac      \\
\hline  DA 55      & 0528$+$134 & 3C 273     & 3C 345     &  2201$+$171    \\
\hline  0215$+$015 & 0716$+$714 & 3C 279     & 1717$+$178 &  CTA 102    \\
\hline  3C 66A     & 0727$-$115 & 1308$+$326 &  OT 081    &  3C 454.3   \\
\hline  0235$+$164 & 0805$-$077 & 1502$+$106 & 1849$+$670 &             \\
\hline  3C 84      & OJ 287     & 1508$-$055 & 1908$-$211 &             \\           
\hline
\end{tabular}
\label{Aller-t1}
\end{center}
\end{table}

\begin{figure*}[t]
\centering
\includegraphics[width=135mm]{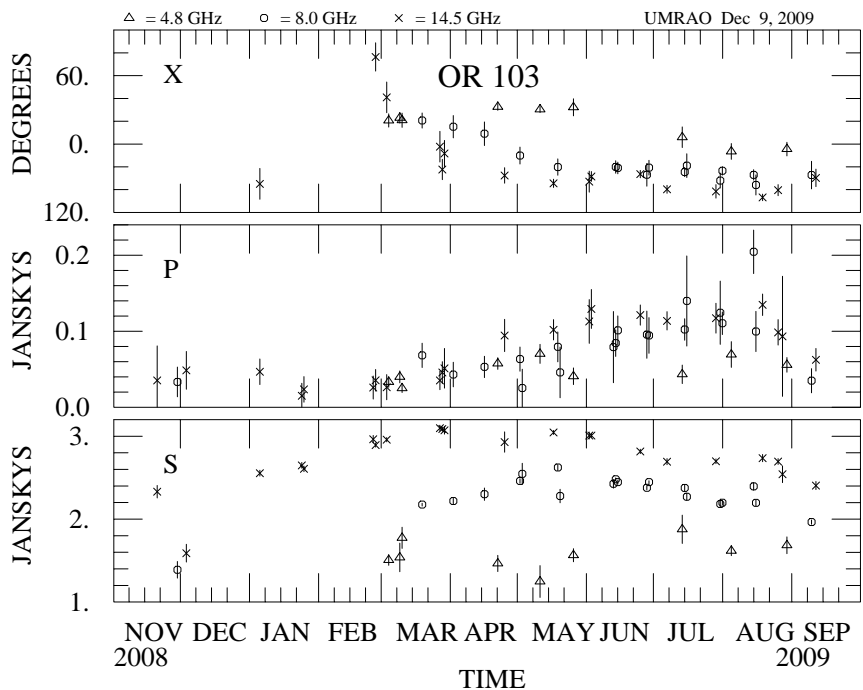}
\caption{From bottom to top, daily averages of total flux density S, polarized flux P, and electric vector
position angle $\chi$ illustrating recent variability
at radio band in the high redshift (z=1.839) QSO 1502+106 (OR 103). Observations at 14.5, 8.0, and 4.8 GHz
are denoted by crosses, circles, and triangles respectively.} \label{Allerpic1-f1}
\end{figure*}

\section{The Source Sample}
Our program sources are listed in Table 1. In addition to blazars, our source sample
includes the radio galaxy 3C 84 (NGC 1275);  $\gamma$-ray emission has been detected in this source
which appears to be associated with compact regions within the radio jet \cite{Abdo09b}. Selection was based on
the following criteria:
1) inclusion in the high confidence list in the LAT catalogue based on the first three
months of {\it Fermi}'s operation \cite{Abdo09}, or identified as
$\gamma$-ray flaring in subsequent Astronomers' Telegrams; 2)  15 GHz total flux density of at least 1 Jy
in early 2009; and 3) membership in the MOJAVE 15 GHz survey \cite{Lister09}. While AGN
light curves regularly show peaks and troughs that can last for months or longer,
we expect these objects to be highly likely to exhibit bright flaring states during cycle 2.
A minimum radio band total flux density of 1 Jy was chosen to ensure good signal-to-noise in the 
multifrequency polarimetry data. Typical fractional linear polarizations in blazars are of order a few percent, 
and the radio band spectra of our targets are characteristically flat or inverted (14.5 GHz total flux density $>$ 4.8 GHz 
total flux density).
The MOJAVE 15 GHz imaging data, typically obtained every few months or longer, will complement our single dish
measurements which are more frequent but which integrate over the entire source area.
About half of our sources are in the Boston University 43 GHz VLBA program, providing
information on morphology and structural changes in the inner jet. 
    
    \begin{figure*}[t]
\centering
\includegraphics[width=135mm]{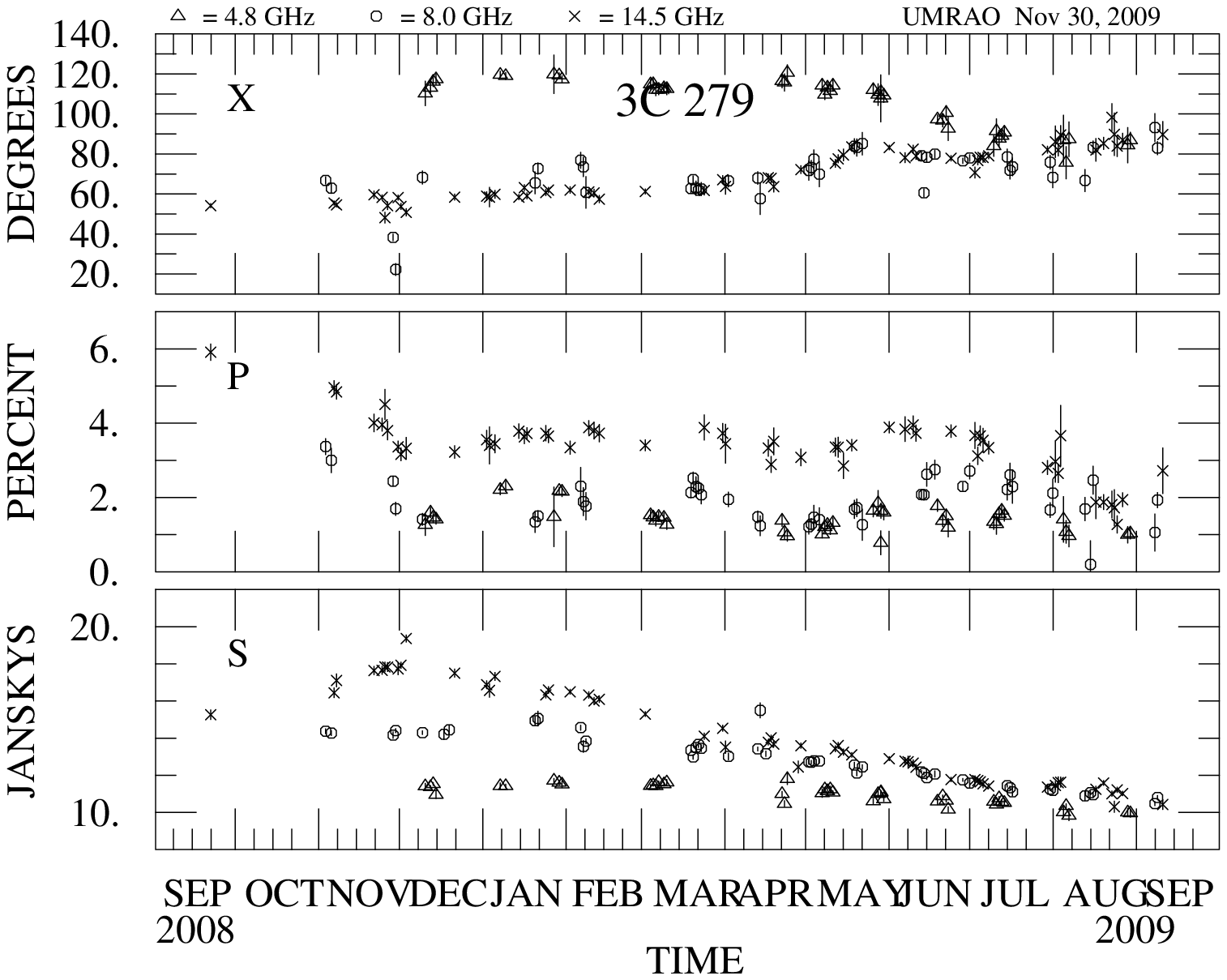}
\caption{From bottom to top, daily averages of total flux density, fractional linear polarization, and
electric vector position angle  at three frequencies illustrating the recent variability
at radio band in 3C 279. Note the changes in polarization in 2009 April-May
which are resolved in these single dish observations. The variations in this source during the mid
1980s were successfully modeled with a transverse shock \cite{Hughes91}.} \label{Allerpic2-f2}
\end{figure*}

  \begin{figure*}[t]
\centering
\includegraphics[width=135mm]{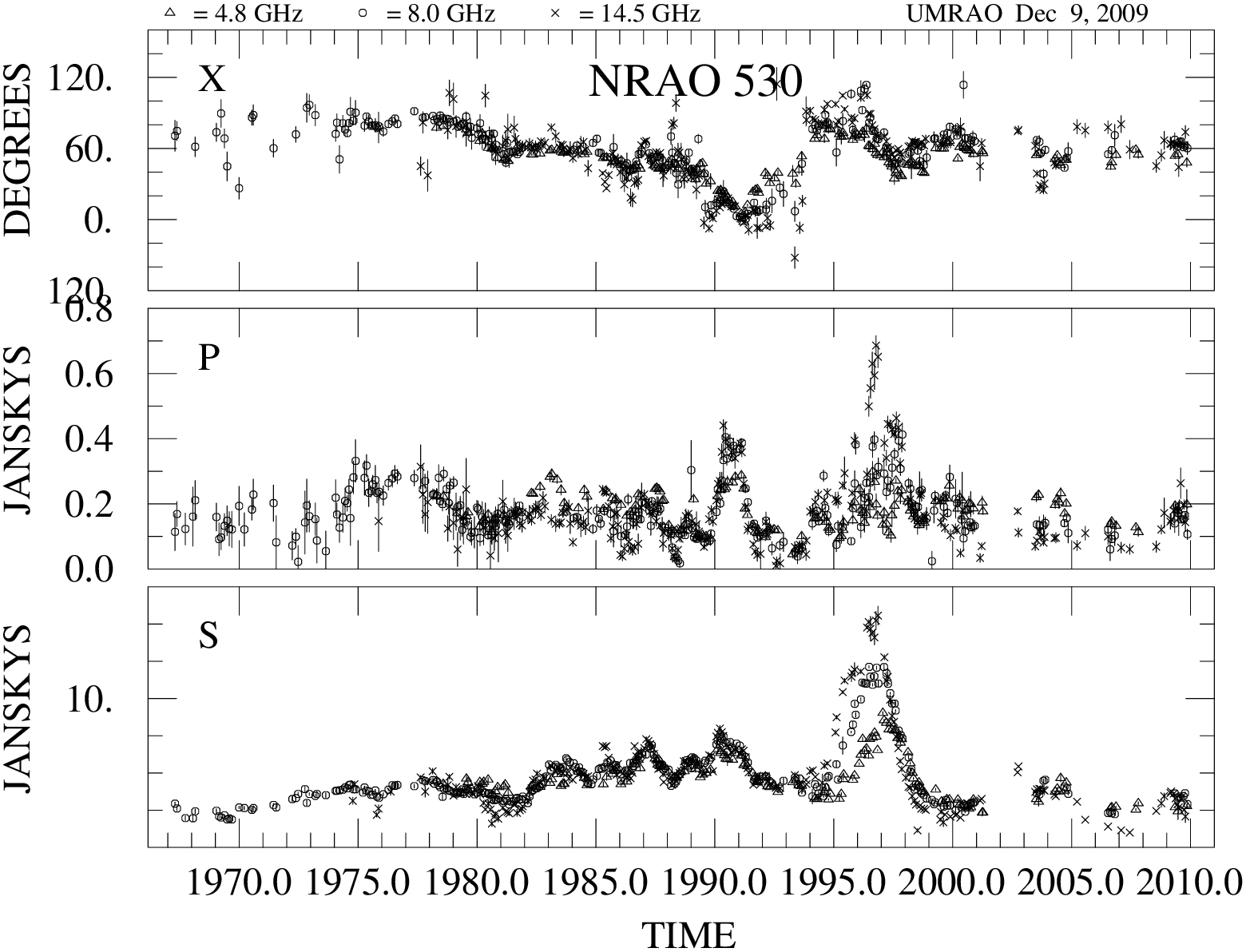}
\caption{From bottom to top, long term monthly averages of the total flux density, polarized flux, and electric vector position
angle in the QSO NRAO 530, a source which was
exceptionally bright in both the $\gamma$-ray and in the radio band during the EGRET era in the 1990s, but which
is currently exhibiting only low level variability in both of these bands.} \label{Allerpic3-f3}
\end{figure*}

As new $\gamma$-ray-flaring radio-bright AGN are identified, they are  added to the UMRAO program if they meet
our criteria for inclusion in the monitoring program; examples of newly added sources are
NRAO~190, a bright EGRET source \cite{McG97} in which recent {\it Fermi}-detected flaring was reported \cite{Cip09}, 
and 3C~345 a well-known bright blazar surprisingly not detected by EGRET.  Additionally, we are observing on a
time-available basis a few sources which were exceptionally bright during the EGRET era but  which are
currently relatively quiescent in both the radio and $\gamma$-ray bands. Included in this group is
NRAO 530 \cite{Bower97}. Several of the sources listed in Table 1 are new to our core program;
 others have been observed by UMRAO for many years.

%



\subsection{Observing Strategy}

Before selecting our sources, we carried out exploratory radio band measurements of a larger group in 2009 March
to evaluate the current radio band variability state and to measure the flux levels for all potential program sources.
Monitoring of our sample commenced
in 2009 August. Throughout cycle 2 we expect to obtain two observations weekly 
at 14.5 GHz (the frequency at which the variability amplitude is highest in AGN), and
 one observation weekly at each of 8.0 and 4.8 GHz for flaring sources. This sampling
rate is matched to the expected variability timescale and duty cycle in the radio band  based on historical
measurements obtained over decades by UMRAO \cite{Aller85}. However, the sampling rate will be increased if warranted,
 e.g. during some phases in the BL Lac object 0716+714
which has historically exhibited large amplitude, very rapid radio band variations.
Each daily measurement consists of a series of on-off measurements over a 30-40 minute
time span; these measurements
are interleaved with observations of a calibrator every 1-2 hours to monitor the antenna gain and pointing,
and to verify the instrumental polarization. As a result of these requirements, 
only 20-25 program sources can be observed in each 24 hour run, and the most active will be selected from
the sample.

\section{Early Program Results}
The first stage of our program, to obtain data exhibiting resolved linear polarization variability
temporally associated with time periods of $\gamma$-ray flaring, is in progress.
During the first two months of the program we have
detected several polarization events, and we are in the process of examining these and our
subsequent data to identify characteristic patterns in the light curves. Resolved flares potentially
suitable for detailed modeling have been identified in 0727-115, 0805-077, 3C 279, 1502+106, and 1510-089.

We are finding that the timescales of the outbursts in polarized flux and periods of ordered temporal
changes in EVPA are relatively short, typically
several weeks to a few months in duration. As an example, we show in Figure~\ref{Allerpic1-f1} the
light curves for 1502+106 (OR~103), a relatively new addition to our core
program, which is highly active at $\gamma$-ray band. The MOJAVE 15 GHz source structure in this source
 is relatively simple, and much of the polarized emission is
associated with a single source region.  A study of $\gamma$-ray flaring in August 2008 
used two MOJAVE images
separated by several months to identify a change in magnetic field orientation \cite{Abdo10},
but the sampling of the imaging data did not permit tracing the change in structure.  During the subsequent
period covered by our monitoring data, there is a resolved outburst in polarized flux
(middle panel) and a systematic swing in EVPA (top panel) which commenced in 2009 February.
 These variations track at 14.5 and
8.0 GHz; but the data follow a different path  at 4.8 GHz. The multifrequency total flux density data 
 (bottom panel) show a spectrum characteristic of a self-absorbed source;
thus the emission at 4.8 GHz most likely arises in a somewhat different physical region of the source, further out from the 
inner region probed by the higher frequencies. In this flare the position angle swing and the
development of the linearly polarized outburst are  characteristic of a shock event. 

 Figure~\ref{Allerpic2-f2} shows results for a structurally complex QSO, 
3C~279, where the contributions from individual core and jet components are blended in the total flux density light curve.
During a series of events in the mid-1980s the variations apparent in the UMRAO
data for this source were successfully modeled assuming a transverse shock \cite{Hughes91}; and
the jet parameters derived in that work will be used as a starting point in the new analysis.

 Some sources exhibiting $\gamma$-ray flaring have not shown large amplitude changes in either
   total flux density or in polarized flux in our observing band (e.g. NRAO 190). We do not 
know yet whether this is due  to frequency-dependent time delays across bands, the masking of variability in the UMRAO 
   source-integrated measurements due to competing, independently evolving source components, or 
   whether more than one scenario is required to explain the origin of the high energy emission.

  Two very bright EGRET sources on the {\it Fermi} LAT list, 0528+134 and NRAO 530, have not
   exhibited high-amplitude flaring in {\it either} the $\gamma$-ray or in the radio band since the launch of {\it Fermi}. We show
the long term light curves for NRAO 530 in Figure~\ref{Allerpic3-f3} which illustrates the relatively low flux level
in the radio band since the launch of {\it Fermi}. The presence of low fluxes in both bands is  consistent with the
expected behavior assuming that the activity is broad band and that the same particles are responsible for the high
and the low energy emission.

%



In the next phase of our work, we will carry out shock modeling of resolved radio band events exhibiting
the shock signature. This will allow us to set limits on the physical conditions in the radio jet during
$\gamma$-ray flaring. A set of resolved radio band flares will be selected for detailed shock-in-jet
modeling following the procedures in our previous analyses \cite{Hughes89,Hughes91} which use multifrequency
linear polarization and total flux density observations as constraints. While our earlier work assumed
a specific shock geometry (transverse shocks), 
the new models will employ a transfer code that admits arbitrary shock orientation, resulting
in extra degrees of freedom which must be constrained. The complementary VLBA imaging data from the
MOJAVE and BU programs will be used to limit the allowable range of orientations. The modeling
is expected to yield information on shock strength, the character of the jet flow, and the low
energy cutoff in the spectrum of the radiating particles,
all of which are  key input parameters in jet emission models designed to explain
the origin of the $\gamma$-ray emission \cite{Dermer09}.

\bigskip 
\begin{acknowledgments}

This work is supported by NASA {\it Fermi} grant NNX09AU16G. The operation of UMRAO
is made possible by funds from the University of Michigan Astronomy Department and by a series of
grants from the NSF, most recently AST-0607523. This research has
made use of data from the MOJAVE database and from the {\it Fermi} website.
\end{acknowledgments}

\bigskip 

\end{document}